# Ptychographic retrieval for complete ultrashort pulse amplitude swing reconstruction

CRISTIAN BARBERO,[1,*] ÍÑIGO J. SOLA,[1,2] AND BENJAMÍN ALONSO[1,2]

[1]*Grupo de investigación en Aplicaciones del Láser y Fotónica (ALF), Universidad de Salamanca, 37008 Salamanca, Spain*
[2]*Unidad de Excelencia en Luz y Materia Estructuradas (LUMES), Universidad de Salamanca, Spain*
*cristianbp@usal.es

**Abstract:** Ultrafast experiments require precise temporal characterization of laser pulses, where pulse reconstruction is typically achieved through iterative retrieval algorithms. In this context, the amplitude swing technique has emerged as a robust and versatile tool for measuring scalar and vector pulses, using a compact inline setup. Here, we develop a unified theory of amplitude swing and introduce a ptychographic iterative engine algorithm to retrieve scalar and vector pulses from the different amplitude swing implementations (conventional and generalizing). For scalar pulses, we show the new capability of retrieving both amplitude and phase. The algorithm is validated with simulated and experimental traces of scalar and vector pulses, which are benefited from the new algorithm. In conclusion, a powerful tool for pulse characterization is provided, extending the capabilities of amplitude swing (e.g., to double pulses) and enhancing the ultrafast optics measurement toolkit.

## 1. Introduction

The temporal measurement of ultrashort laser pulses is key for every application of these unique light sources. To do this, some established techniques are FROG [1], SPIDER [2], MIIPS [3] or d-scan [4,5]. More recently, new techniques have been introduced to expand the existing capacities (as discussed below), for example TIPTOE [6], FROSt [7], or amplitude swing (a-swing) [8]. These techniques are based on the reconstruction of the pulse from a measured signal.

Regarding the reconstruction, ptychographic iterative engine (PIE) algorithms have been proven to be very successful in different areas. In general terms, they are related to measurements in which two properties (e.g., illumination or objects) are relatively scanned while a signal (trace or pattern) is recorded. Originally, ptychography was used in microscopy [9,10]. Related to pulse measurement, ptychography has been applied in the last few years in ultrafast pulse reconstruction [11], attosecond pulses measurement [12], XFROG retrieval [13], time-domain ptychography for single-cycle pulses [14] or with phase-only transfer functions [15], and FROG [16,17] or FROSt [7] reconstructions, among others [18]. In d-scan retrievals, they have been used projections-based retrieval algorithms [19], ptychographic retrievals [20], or ptychographic retrieval to evaluate pulse incoherence [21]. Also, it has been applied to spatiotemporal pulse measurements [22,23] and nanosecond scale imaging [24].

In the present work, we deal with the a-swing technique [8], which is a recent strategy to measure ultrashort pulses. In the original set-up of this technique, a rotating multi-order waveplate (MWP) followed by a linear polarizer (LP) is used to create in a bulk configuration two delayed pulse replicas with varying relative amplitude. The interfering replicas produce a second harmonic generation signal (SHG) whose spectrum is measured as a function of the MWP orientation, giving the 2D a-swing trace. In the fundamentals of the technique [8] it is shown that the experimental a-swing setup is simple, compact, and robust to be applied under different scenarios [25]. The delay between replicas introduced by the MWP must be of the order of the Fourier limit of the pulse to be measured (with a remarkable flexibility up to a ratio of 3, up or down), while highly chirped unknown pulses can be measured with the same waveplate [8,25]. This flexibility has been shown to be beneficial e.g., in the measurement of pulses from the visible (620 nm) to the near-infrared (1550 nm) with the same a-swing [26], its integration into more complex (e.g., spatiotemporal [27]) detections, and in the measurement of few-cycle pulses [28]. Recently, in the generalizing a-swing [29] that we will recall in this work, it has been shown that a complex amplitude modulation can be advantageous under certain practical circumstances. Newly, it has been shown that a-swing can measure vector pulses from a single trace [30] (as opposed to common vector techniques [31–33]) using the conventional configuration in [8].

To reconstruct the unknown pulse from the a-swing trace, iterative algorithms based on nonlinear optimization (for example, with the Levenberg-Marquardt (LM) algorithm) have been used both for scalar [8] and vector [30] pulses, while differential evolution algorithms have been used too for scalar pulses [26]. Those strategies can reconstruct the a-swing measurement, while they are partial since the number of unknowns is limited in practice. They rely on a particular parametrization of the phase and can retrieve the spectral phase only, whereas the spectral amplitude must be measured. This limitation means that each retrieval takes few minutes. Also, the vector pulse retrieval [30] increases in complexity and needs several stages to complete a retrieval, which are valid for a conventional a-swing trace but could not be extended to a generalizing a-swing trace. Here, we show that the ptychographic algorithm developed can fix those issues.

Comparing to PIE algorithms applied to, e.g., microscopy or FROG, where the illumination/object and the pulse are constant during the scan, in a-swing the replicas are modulated during the scan, so that it is to be specially considered here in the introduced algorithm. Another challenge that we will solve below is that only the SHG from one polarization projection is measured.

In this work, we firstly develop a unified theory of the conventional and generalizing a-swing both for scalar and vector pulses, for which we introduced the corresponding Jones matrices and the inverse propagation. We use them to design a ptychographic reconstruction algorithm, which is faster as the unknown field is updated for every slice of the trace and not after evaluation of the entire trace. Apart from the improvement in vector pulse reconstructions, here we show the application to retrieve both the amplitude and phase of scalar pulses, and to reconstruct double pulses, among others.

## 2. Materials and methods

### A. Amplitude swing theory

Here, we unify the formalism and description for conventional a-swing for scalar [8] and vector pulses [30], and the recently introduced generalizing a-swing [29], which is upgraded here for vector pulse measurements. This formalism is based on Jones calculus that is applied to completely polarized ultrashort pulsed lasers. In conventional a-swing [8], the input pulse propagates through the calibrated MWP and the LP at 0° (Fig. 1(a)). The MWP angle $\theta_M$ (fast axis with respect to the *x*-axis) is scanned to create the pulse replicas with relative amplitude variation after the LP. If the input pulse is LP (we use at 0° by convention), two amplitude modulated replicas interfere after the LP [8], whereas for input vector pulses there are 4 interfering pulses (coming from the replicas of each polarization component) [30]. In the case of generalizing a-swing [29], the input pulse propagates through a zero-order waveplate (ZWP) at an angle $\theta_Z$ (fast axis) that is scanned, followed by a static MWP (at 45°), before the LP at 0° (Fig. 1(b)). The resulting effect is a complex amplitude modulation of the pulse replica that depends on the phase retardation of the ZWP [29]. After the LP at 0°, the modulated interfering replicas are up-converted in a SHG crystal and the SHG spectra are recorded as a function of the scanning angle, in a 2D a-swing trace, from which the input pulse is reconstructed afterwards [8,29,30].

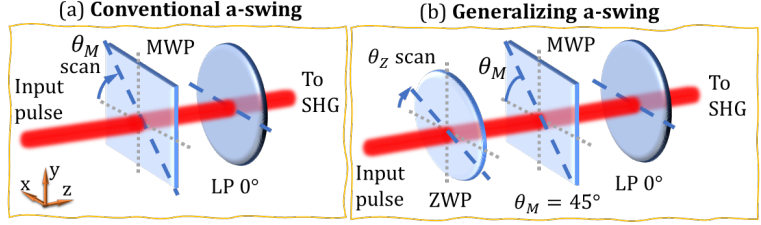

Fig. 1. Scheme of the shaping used for the pulse replica modulation in the experimental setup for (a) conventional and (b) generalizing a-swing. (a) A rotating MWP creates the pulse replicas. (b) A rotating ZWP followed by a static MWP (45°) creates the pulse replicas. The replicas are projected with a linear polarizer (LP) at 0° so that they interfere. After that, the SHG is generated in a nonlinear crystal, the remaining fundamental signal is optionally removed with a filter and the SHG is measured with a spectrometer.

Whatever configuration used, the a-swing modulation can be modeled as a transfer function given by a 2×2 matrix, $\mathbf{H}(\omega, \theta)$, that stands for the waveplates propagation (rotating MWP versus rotating ZWP and static MWP, for conventional and generalizing configurations, respectively), and depends on the angular frequency $\omega$ and the waveplates calibration and orientation ($\theta$ denotes $\theta_M$ or $\theta_Z$ for conventional and generalizing configurations, respectively). This transfer matrix is applied to the input pulse, $\mathbf{E}^{(\text{in})}(\omega)$, which is in general a vector pulse (in the case of scalar pulses, we simply set $E_y^{(\text{in})} = 0$). Note that we keep the linear polarizer out of the propagation matrix since otherwise it would not be invertible, which would mean here that the pulse could not be back propagated in the retrieval (the implications in the algorithm will be discussed later). The operator $\mathbf{P}_x = (1 \quad 0)$ takes the (scalar) $x$-projection of the pulse (i.e., the LP at 0°). Then the expression for the SHG spectrum corresponding to the a-swing trace, $S_{\text{swi}}^{(\text{exp})}(\omega, \theta)$, can be given in compact form as

$$S_{\text{swi}}^{(\text{exp})}(\omega, \theta) = \left|\mathcal{F}^{-1}\left[\left\{\mathcal{F}\left[\mathbf{P}_x \cdot \mathbf{H}(\omega', \theta) \cdot \mathbf{E}^{(\text{in})}(\omega')\right]\right\}^2\right]\right|^2 \tag{1}$$

where $\mathcal{F}$ denotes Fourier-transform.

From Jones calculus, we calculate the $\mathbf{H}$ matrix expressed in the $xy$ axes for conventional a-swing [8,30]

$$\mathbf{H_C}(\omega, \theta) = \begin{pmatrix} H_{xx} & H_{xy} \\ H_{yx} & H_{yy} \end{pmatrix} = \begin{pmatrix} \cos^2\theta e^{i\phi_f} + \sin^2\theta e^{i\phi_s} & \cos\theta\sin\theta(e^{i\phi_f} - e^{i\phi_s}) \\ \cos\theta\sin\theta(e^{i\phi_f} - e^{i\phi_s}) & \sin^2\theta e^{i\phi_f} + \cos^2\theta e^{i\phi_s} \end{pmatrix} \tag{2}$$

and for generalizing a-swing

$$\mathbf{H_G}(\omega, \theta) = \begin{pmatrix} H_{xx} & H_{xy} \\ H_{yx} & H_{yy} \end{pmatrix} = \begin{pmatrix} G_f e^{i\phi_f} + G_s e^{i\phi_s} & e^{i\phi_r} G_s^* e^{i\phi_f} - e^{i\phi_r} G_f^* e^{i\phi_s} \\ G_f e^{i\phi_f} - G_s e^{i\phi_s} & e^{i\phi_r} G_s^* e^{i\phi_f} + e^{i\phi_r} G_f^* e^{i\phi_s} \end{pmatrix} \tag{3}$$

where we recall the definitions of the modulation factors [29]

$$\begin{aligned} G_f &= \frac{1}{4}\left[(1 + \cos 2\theta + \sin 2\theta) + (1 - \cos 2\theta - \sin 2\theta)e^{i\phi_r}\right] \\ G_s &= \frac{1}{4}\left[(1 + \cos 2\theta - \sin 2\theta) + (1 - \cos 2\theta + \sin 2\theta)e^{i\phi_r}\right] \end{aligned} \tag{4}$$

In those equations, $\phi_f(\omega)$ and $\phi_s(\omega)$ are the spectral phases (dispersion) introduced by the fast and slow axes of the calibrated MWP, and $\phi_r$ is the constant phase retardation of the ZWP (only applies to generalizing a-swing).

It is also necessary to define the inverse transfer matrix, $\mathbf{H}^{-1}$, for later use for back-propagation during the pulse reconstruction

$$\mathbf{H}^{-1}(\omega, \theta) = \frac{1}{\det(\mathbf{H})}\text{adj}(\mathbf{H}) = \frac{1}{\det(\mathbf{H})}\begin{pmatrix} H_{yy} & -H_{xy} \\ -H_{yx} & H_{xx} \end{pmatrix} \tag{5}$$

where adj and det, denote the adjugate and the determinant of **H**, being $\det(\mathbf{H_C}) = e^{i\phi_f}e^{i\phi_s}$ and $\det(\mathbf{H_G}) = e^{i\phi_r}e^{i\phi_f}e^{i\phi_s}$, for conventional and generalizing a-swing, respectively. Note that dividing by $\det(\mathbf{H})$ is essential to account for back propagation thus taking out the dispersion of the waveplates (MWP and ZWP).

## B. Ptychographic algorithm

Once the pulse forward- and backward-propagation have been established, we develop the ptychographic algorithm for the a-swing trace reconstruction. We present the general case of vector pulse retrieval, being directly valid for scalar pulses too. Firstly, let's consider an initial guess for the unknown pulse, $\mathbf{E}^{(in)}(\omega)$. An angle $\theta$ is selected to apply the reconstruction algorithm. At every step, we choose a random $\theta = \theta_j$ (with $j = 1, \dots, N_\theta$, representing the orientations measured in the MWP or ZWP scan) until the entire a-swing trace is used. Then, the process is repeated in the subsequent step. The forward (fwd) propagated pulse for $\theta = \theta_j$ is

$$\mathbf{E}^{(fwd)}(\omega, \theta) = \mathbf{H}(\omega, \theta) \cdot \mathbf{E}^{(in)}(\omega) \quad (6)$$

Since we set the LP at 0° to select the interfering replicas, the a-swing fundamental field is $E_\theta(t) = \mathcal{F}^{-1}\left[E_x^{(fwd)}(\omega, \theta)\right]$. The SHG signal of the field in the temporal domain is given by

$$\psi_\theta(t) = \left(E_\theta(t)\right)^2 \quad (7)$$

Then, we transform it into the spectral domain, where we do the replacement of the measured amplitude obtained from the SHG spectra of the experimental a-swing trace, $S_{swi}^{(exp)}(\omega, \theta)$, in this way

$$\psi'_\theta(t) = \mathcal{F}^{-1}\left\{\sqrt{S_{swi}^{(exp)}(\omega, \theta)} \frac{\mathcal{F}[\psi_\theta(t)]}{|\mathcal{F}[\psi_\theta(t)]|}\right\} \quad (8)$$

Then, we can apply the update equation for the fundamental a-swing field

$$E'_\theta(t) = E_\theta(t) + \beta \frac{E_\theta^*(t)}{|E_\theta(t)|^2 + \alpha \cdot \max(|E_\theta(t)|^2)} \left(\psi'_\theta(t) - \psi_\theta(t)\right) \quad (9)$$

where $\beta$ indicates the update velocity and $\alpha$ avoids dividing by 0. In the literature, it is used $0 < \beta < 1$ and $0 < \alpha < 1$, here we find a good performance with $\beta = 0.5$ and $\alpha = 0.15$. Note that some previous works divide by a constant value $\max(|E_\theta(t)|^2)$ in Eq. (9), while here we find that the reconstruction is better when keeping the field intensity $|E_\theta(t)|^2$ with the $\alpha$ correction.

From that, we can calculate the updated (upd) a-swing field in the spectral domain $E_x^{(upd)}(\omega, \theta) = \mathcal{F}[E'_\theta(t)]$. To retrieve the input pulse, it is necessary to back propagate with $\mathbf{H}^{-1}$. As said before, in principle the LP at 0° makes the back propagation non-invertible since $E_y^{(fwd)}(\omega)$ was filtered out before the SHG. To overcome it, the strategy followed is to save the y-component of the forward propagation field $E_y^{(fwd)}(\omega)$ (see Eq. (6)), and use it here (without update) in the following expression

$$\mathbf{E}^{(pty)}(\omega) = \begin{pmatrix} E_x^{(pty)}(\omega) \\ E_y^{(pty)}(\omega) \end{pmatrix} = \mathbf{H}^{-1}(\omega, \theta) \cdot \begin{pmatrix} E_x^{(upd)}(\omega) \\ E_y^{(fwd)}(\omega) \end{pmatrix} \quad (10)$$

In this way, we obtain the ptychographic (pty) reconstructed pulse, $\mathbf{E}^{(pty)}(\omega)$, at a certain iteration. This pulse is used as the input pulse $\mathbf{E}^{(in)}(\omega)$ in the next iteration, so that the process described in Eq. (6-10) is repeated until a preset convergence criterion is met or maximum number of iterations is reached. Note that in the case of scalar pulse reconstruction, $E_y^{(pty)}(\omega)$ is set to 0 (in fact, it can be checked when convergence is reached).

Optionally, if the input pulse spectrum is measured, the retrieved field in Eq. (10) can be constrained to the experimental amplitude while keeping the spectral phase obtained from the ptychographic algorithm.

## 3. Results

We first validate the PIE algorithm by retrieving simulated scalar and vector pulses, both with the conventional (rotating MWP) and generalizing (here using a rotating HWP or QWP) a-swing. Then, we reconstruct experimental scalar and vector pulses, using conventional and generalizing (HWP) a-swing measurements.

### A. Application to simulated scalar pulses

To validate the PIE algorithm, we first retrieve different simulated scalar pulses for the three mentioned a-swing configurations (Fig. 2), retrieving both amplitude and phase. In the first row, we simulate and retrieve a pulse with a gaussian spectrum centered at 800 nm, a Fourier-limited (FL) pulse duration $t_{FL} = 50$ fs (full-width at half maximum, FWHM), and a group delay dispersion (GDD) and third order dispersion (TOD) of 1800 fs$^2$ and -90000 fs$^3$, respectively, plus a phase oscillatory term. In the second row, we simulate a pulse in the few-cycle regime, being $t_{FL} = 5$ fs, with a pure oscillatory spectral phase, leading to a complex temporal structure. In the third row, we simulate a double pulse with 200-fs pulse separation, each one with $t_{FL} = 25$ fs, GDD = 400 fs$^2$, and TOD = 1500 fs$^3$. The MWP used in these simulations introduces a delay equal to the FL pulse duration, and a $0.34\pi$ phase retardation (for 800 nm). We have retrieved the three pulses with all configurations (in Fig. 2 we show 3 different configurations as an example), and in all cases the retrieved pulse matches the simulated pulse. The guess pulse is defined by a flat spectrum and a random phase. The phase converges quickly, in ~20 seconds, while the spectrum initially exhibits an artificial modulation that may require more iterations to vanish completely. Note that in d-scan it has been reported that some oscillations appear in the retrieved spectral amplitude and increase for a higher number of iterations [20].

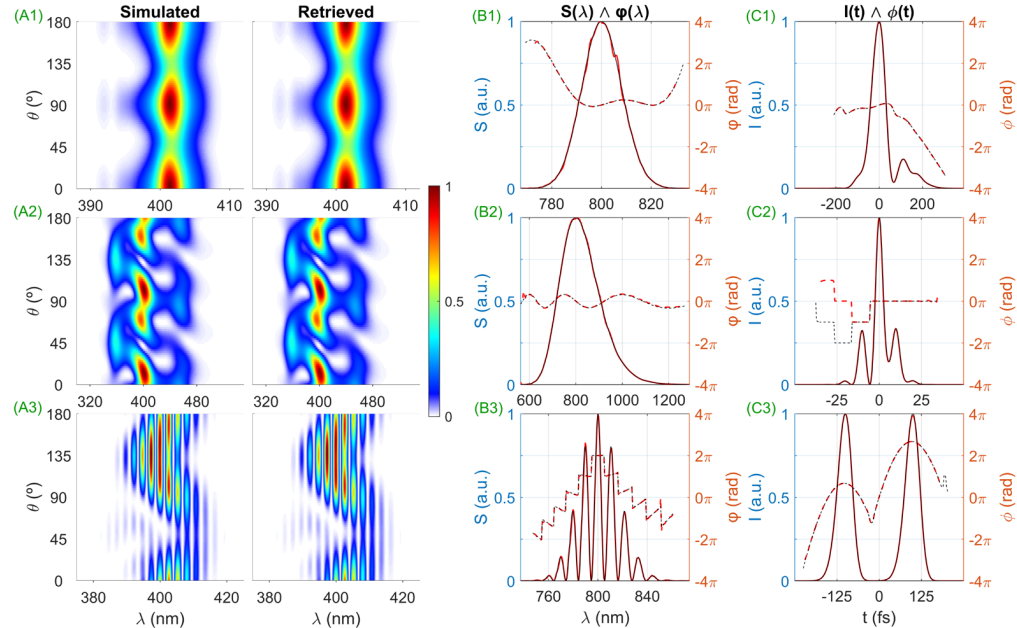

Fig. 2. Reconstruction of simulated scalar pulses from a-swing traces (A); simulated (black) and retrieved (red) intensity (solid) and phase (dashed), in the spectral (B) and temporal (C) domain. Row 1: multi-cycle pulse retrieved using conventional a-swing. Row 2: few-cycle pulse retrieved from a HWP generalizing trace. Row 3: retrieval of a double pulse using the QWP generalizing version.

## B. Application to simulated vector pulses retrieval

Vector pulses, whose polarization state is time-dependent, are characterized by the spectral amplitude and phase of two components, including the relative phase between them. To extract this information, in [30] is used a multi-step reconstruction strategy to gradually gain and refine the phase retrieval, which takes advantage of the fact that the conventional trace does not depend on $E_y(\omega)$ at $\theta = 0°, 90°, 180°$. Since the generalizing a-swing trace depends on both pulse components along the entire trace, this strategy is not valid for such amplitude modulation. This new algorithm allows a simultaneous reconstruction of both phases (the amplitudes are not retrieved in this case), extending the applicability to all a-swing configurations, while highly increasing the retrieval speed, from several minutes to approximately one minute.

Fig. 3 shows the retrieval of the horizontal and vertical polarization projections of different vector pulses, using a MWP defined by a 50-fs delay and a $0.34\pi$ phase retardation for 800 nm. In the first row, we simulate a vector pulse in which the spectra are modulated (as occurs to the projections of a pulse propagating through a retarder plate due to the interference between the fast and slow replicas), thus the spectral phase presents multiple $\pi$ jumps, and is retrieved from the conventional a-swing trace. This case is challenging for the state-of-the-art algorithms [8,30], in which the phase is parametrized by its derivative values in 32 points of the frequency axis, restricted to phases with limited abrupt changes. In the second row, a twisting polarization pulse, i.e., linearly polarized with time-varying azimuth, is retrieved using the HWP generalizing a-swing. This pulse is the superposition of two counter-rotating circularly polarized pulses with a GDD of ±2000 fs$^2$. Experimentally, it can be generated by propagating slightly spectrally sheared chirped perpendicular polarization components through two consecutive QWP (multi- and zero-order) oriented at 45° [34]. In the third row, using the QWP generalizing version, a pulse with gaussian spectra is retrieved, with projections centered at 797 and 803 nm, and oscillatory spectral phases plus a relative delay of 35 fs. All retrievals show excellent agreement with the simulated pulses.

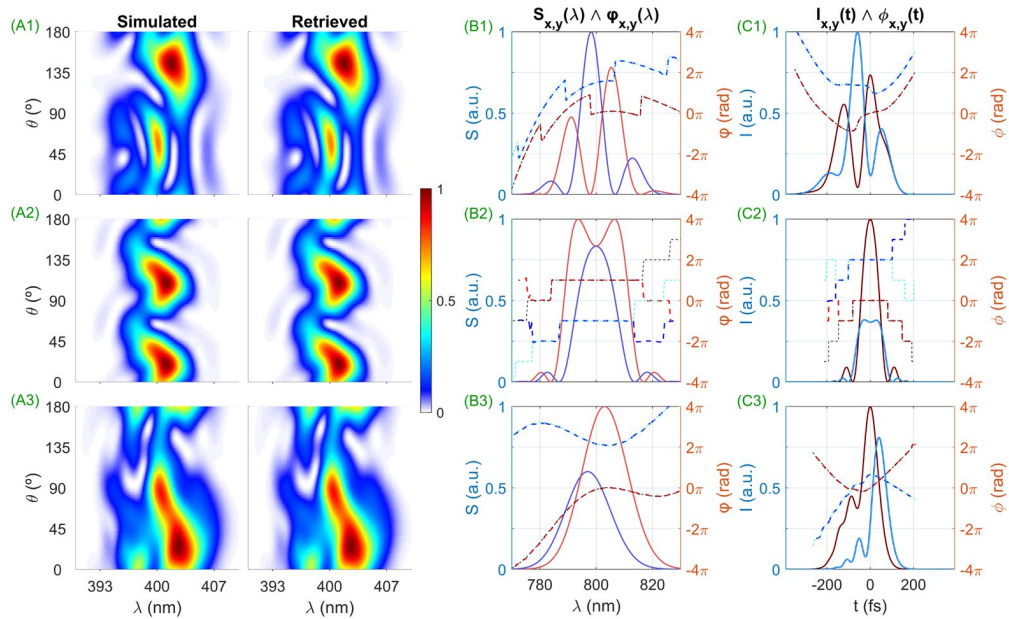

Fig. 3. Reconstructions of simulated vector pulses using a-swing traces: (A) simulated and retrieved traces; (B) spectra (solid) and spectral phases (dashed) of simulated (black and cyan) and retrieved (red and blue) vector pulse components; (C) simulated (black and cyan) and retrieved (red and blue) temporal intensities (solid) and phases (dashed). Row 1: spectral phases with pi jumps, retrieved from the conventional a-swing. Row 2: retrieval of a twisting polarization pulse using the HWP generalizing version. Row 3: retrieval of a pulse with oscillatory spectral phases with the QWP generalizing a-swing.

## C. Application to experimental measurement of scalar pulses

Here we apply the developed PIE algorithm to experimental a-swing measurements. The pulses are emitted by a Ti:sapphire CPA laser system (Spitfire Ace, Spectra Physics), with a central wavelength of ~800 nm and a FL pulse duration of ~60 fs (FWHM). By adjusting the compressor of the laser system, we measure two pulses with positive and negative GDD, using the conventional (rotating 2-mm quartz MWP, 65 fs delay and $0.34\pi$ phase retardation for 800 nm) and generalizing (rotating HWP followed by static 2-mm quartz MWP) setups (Fig. 1), respectively. We use the measured fundamental spectrum and retrieve the spectral phase with the developed PIE, which takes ~30 s, and the existing LM algorithms (~3 min), observing a high agreement between both retrieved pulses for the two a-swing configurations (Fig. 4).

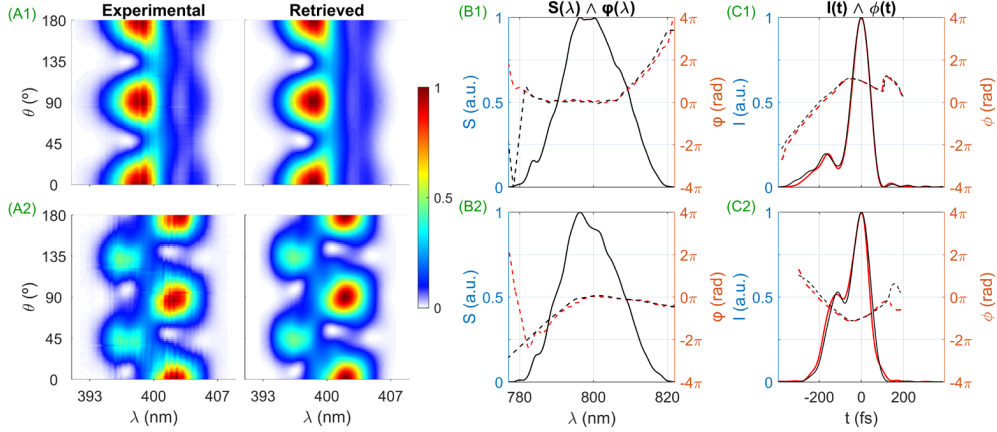

Fig. 4. Application of the PIE algorithm to experimental measurements of scalar pulses: (A) experimental and retrieved conventional (row 1) and HWP generalizing (row 2) traces; (B) measured spectrum (solid grey) and retrieved spectral phase (dashed); (C) retrieved temporal intensity (solid) and phase (dashed), using PIE (black) and LM (red) algorithms.

## D. Application to experimental measurement of vector pulses

Finally, we apply PIE to experimental a-swing traces of vector pulses. These pulses are generated by propagating the previous scalar pulses through a vector shaping system, which consists of a retarder plate oriented at 30° with respect to the scalar pulse polarization (as done in [30]), introducing a 65 fs delay between the fast and slow replicas. After generation, the vector pulse is sent to the a-swing detection system, both conventional and generalizing (Fig. 1), using the same rotating MWP and HWP as in the scalar case of Section 3.C. Fig. 5 shows the reconstruction of said generated vector pulse from the conventional and generalizing traces. We validate both retrievals by comparing them with the reconstructed pulse from the conventional trace, using the multi-step optimization strategy based on the LM algorithm, since this strategy is not valid for the generalizing trace [30] as said before. The agreement is good for both implementations, providing either a novel and faster tool to fully retrieve these complex pulses, depending on whether generalizing or conventional approach is used. Since the phase is not parametrized, the PIE can retrieve more complex phases than the multi-step strategy.

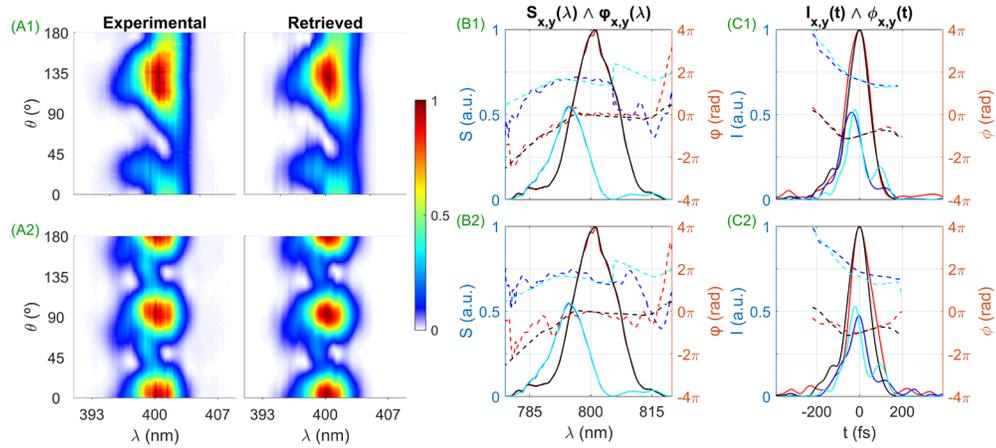

Fig. 5. Experimental vector pulse reconstruction: (A) measured and simulated conventional (row 1) and HWP generalizing (row 2) a-swing traces; (B) measured spectra (solid) and retrieved spectral phases (dashed); (C) retrieved temporal intensities (solid) and phases (dashed), using PIE (red and blue) and LM (black and cyan) algorithms.

## 4.  Discussion and Conclusions

In sum, we have developed a ptychographic engine algorithm for the retrieval of ultrashort scalar and vector pulses from conventional and generalizing amplitude swing measurements. First, we have presented a unified formalism that encompasses all the cases regarding the pulse to be measured and the kind of amplitude modulation. Using this theory, we have designed the ptychographic reconstruction algorithm. Initially, it is validated by retrieving multiple simulated scalar and vector pulses, which are of interest in many optical applications. Finally, it is successfully applied to experimental measurements of scalar and vector pulses.

This algorithm evaluates the trace slice by slice, which implies important advantages over the previously demonstrated Levenberg-Marquardt and differential evolution algorithms. First, the pulse phase does not need to be parametrized, which can be restrictive, but it is an arbitrary function without a priori constraints. This allows the retrieval of phases with abrupter changes, e.g., double pulses (shown here for the first time for a-swing) or vector pulses generated by a retarder plate, which exhibits π jumps when the spectrum is null. Another important advantage is that, in the case of scalar pulses, it can simultaneously retrieve the pulse amplitude and phase (the amplitude can be optionally constrained to the measured spectrum). In the case of vector pulses, it is valid for any applied amplitude modulation, including the generalizing, which was not possible to retrieve up to date. Furthermore, the speed is increased from ~2 minutes to ~20 seconds and from 10-15 minutes to ~1 minute, for scalar and vector reconstructions, respectively. Another interesting capability is the robustness against energy fluctuations during the measurement scan (present, for example, in high-intensity lasers and/or with pointing instability), since the trace is reconstructed slice by slice, instead of using the entire trace.

In conclusion, the retrieval algorithm presented here contributes to pulse characterization by allowing a faster and robust reconstruction of scalar and vector pulses from conventional and generalizing amplitude swing traces. These traces can be measured with simple and compact setups, easily adaptable to different pulse parameters, as spectral bandwidth, chirp, or central wavelength. All these characteristics, in conjunction with the developed retrieval strategy, make the whole measurement process —setup preparation, data acquisition and pulse reconstruction— simple, fast, robust, and versatile.

**Funding.** European Regional Development Fund and Consejería de Educación, Junta de Castilla y León (SA108P24); Ministerio de Ciencia e Innovación (PID2020-119818GB-I00, PID2023-149836NB-I00).

**Acknowledgment.** C. Barbero acknowledges the support from Consejería de Educación of Junta de Castilla y León and Fondo Social Europeo Plus, Programa Operativo de Castilla y León, through their Ph.D. grant program.

**Disclosures.** IS: Universidad de Salamanca (P), BA: Universidad de Salamanca (P).

**Data availability.** Data underlying the results presented in this paper are not publicly available at this time but may be obtained from the authors upon reasonable request.

**References**

1. D. J. Kane and R. Trebino, "Characterization of Arbitrary Femtosecond Pulses Using Frequency-Resolved Optical Gating," IEEE J Quantum Electron **29**, 571–579 (1993).
2. C. Iaconis and I. A. Walmsley, "Spectral phase interferometry for direct electric-field reconstruction of ultrashort optical pulses," Opt Lett **23**, 792–794 (1998).
3. V. V Lozovoy, I. Pastirk, and M. Dantus, "Multiphoton intrapulse interference IV Ultrashort laser pulse spectral phase characterization and compensation," Opt Lett **29**, 775–777 (2004).
4. M. Miranda, C. L. Arnold, T. Fordell, F. Silva, B. Alonso, R. Weigand, A. L'Huillier, and H. Crespo, "Characterization of broadband few-cycle laser pulses with the d-scan technique," Opt Express **20**, 18732–18743 (2012).
5. I. Sytcevich, C. Guo, S. Mikaelsson, J. Vogelsang, A.-L. Viotti, B. Alonso, R. Romero, P. T. Guerreiro, Í. J. Sola, A. L'Huillier, H. Crespo, M. Miranda, and C. L. Arnold, "Characterizing ultrashort laser pulses with second harmonic dispersion scans," Journal of the Optical Society of America B **38**, 1546–1555 (2021).
6. S. B. Park, K. Kim, W. Cho, S. I. Hwang, I. Ivanov, C. H. Nam, and K. T. Kim, "Direct sampling of a light wave in air," Optica **5**, 402–408 (2018).
7. A. Leblanc, P. Lassonde, S. Petit, J.-C. Delagnes, E. Haddad, G. Ernotte, M. R. Bionta, V. Gruson, B. E. Schmidt, H. Ibrahim, E. Cormier, and F. Légaré, "Phase-matching-free pulse retrieval based on transient absorption in solids," Opt Express **27**, 28998–29015 (2019).
8. B. Alonso, W. Holgado, and Í. J. Sola, "Compact in-line temporal measurement of laser pulses with amplitude swing," Opt Express **28**, 15625–15640 (2020).
9. J. M. Rodenburg and H. M. L. Faulkner, "A phase retrieval algorithm for shifting illumination," Appl Phys Lett **85**, 4795–4797 (2004).
10. A. M. Maiden and J. M. Rodenburg, "An improved ptychographical phase retrieval algorithm for diffractive imaging," Ultramicroscopy **109**, 1256–1262 (2009).
11. D. Spangenberg, E. Rohwer, M. H. Brügmann, and T. Feurer, "Ptychographic ultrafast pulse reconstruction," Opt Lett **40**, 1002–1005 (2015).
12. M. Lucchini, M. H. Brügmann, A. Ludwig, L. Gallmann, U. Keller, and T. Feurer, "Ptychographic reconstruction of attosecond pulses," Opt Express **23**, 29502–29513 (2015).
13. A. M. Heidt, D.-M. Spangenberg, M. Brügmann, E. G. Rohwer, and T. Feurer, "Improved retrieval of complex supercontinuum pulses from XFROG traces using a ptychographic algorithm," Opt Lett **41**, 4903–4906 (2016).
14. T. Witting, D. Greening, D. Walke, P. Matia-Hernando, T. Barillot, J. P. Marangos, and J. W. G. Tisch, "Time-domain ptychography of over-octave-spanning laser pulses in the single-cycle regime," Opt Lett **41**, 4218–4221 (2016).
15. D.-M. Spangenberg, M. Brügmann, E. Rohwer, and T. Feurer, "Extending time-domain ptychography to generalized phase-only transfer functions," Opt Lett **45**, 300–303 (2020).
16. P. Sidorenko, O. Lahav, Z. Avnat, and O. Cohen, "Ptychographic reconstruction algorithm for frequency-resolved optical gating: super-resolution and supreme robustness," Optica **3**, 1320 (2016).
17. D. J. Kane, "Comparison of the Ptychographic Inversion Engine to Principal Components Generalized Projections," IEEE Journal of Selected Topics in Quantum Electronics **25**, 8900408 (2019).
18. D. J. Kane and A. B. Vakhtin, "A review of ptychographic techniques for ultrashort pulse measurement," Prog Quantum Electron **81**, 100364 (2022).
19. M. Miranda, J. Penedones, C. Guo, A. Harth, M. Louisy, L. Neoričić, A. L'Huillier, and C. L. Arnold, "Fast iterative retrieval algorithm for ultrashort pulse characterization using dispersion scans," Journal of the Optical Society of America B **34**, 190–197 (2017).
20. T. Gomes, M. Canhota, and H. Crespo, "Temporal characterization of broadband femtosecond laser pulses by surface third-harmonic dispersion scan with ptychographic retrieval," Opt Lett **47**, 3660–3663 (2022).
21. A. M. Wilhelm, C. G. Durfee, D. E. Adams, and D. D. Schmidt, "Multi-mode root preserving ptychographic phase retrieval algorithm for dispersion scan," Opt Express **29**, 22080–22095 (2021).
22. D. Goldberger, J. Barolak, D. Schmidt, B. Ivanic, C. A. M. Schrama, C. Car, R. Larsen, C. G. Durfee, and D. E. Adams, "Single-pulse, reference-free, spatiotemporal characterization of ultrafast laser pulse beams via broadband ptychography," Opt Lett **48**, 3455 (2023).

23. D. Cruz-Delgado, A. Perez-Leija, N. K. Fontaine, D. N. Christodoulides, M. A. Bandres, and R. Amezcua-Correa, "Ptychography for Multidimensional Characterization of Spatiotemporal Ultrafast Pulses," ACS Photonics **11**, 18–23 (2023).
24. A. Veler, M. Birk, C. Dobias, R. A. Correa, P. Sidorenko, and O. Cohen, "Single-shot ptychographic imaging of non-repetitive ultrafast events," Opt Lett **49**, 178–181 (2024).
25. Í. J. Sola and B. Alonso, "Robustness and capabilities of ultrashort laser pulses characterization with amplitude swing," Sci Rep **10**, 18364 (2020).
26. M. López-Ripa, Í. J. Sola, and B. Alonso, "Amplitude swing ultrashort pulse characterization across visible to near-infrared," Opt Laser Technol **164**, 109492 (2023).
27. M. López-Ripa, Í. J. Sola, and B. Alonso, "Bulk lateral shearing interferometry for spatiotemporal study of time-varying ultrashort optical vortices," Photonics Res **10**, 922–931 (2022).
28. M. López-Ripa, Ó. Pérez-Benito, B. Alonso, R. Weigand, and Í. Sola, "Few-cycle pulse retrieval using amplitude swing technique," Opt Express **32**, 21149–21159 (2024).
29. M. López-Ripa, Í. J. Sola, and B. Alonso, "Generalizing amplitude swing modulation for versatile ultrashort pulse measurement," Opt Express **31**, 34428–34442 (2023).
30. C. Barbero, B. Alonso, and Í. J. Sola, "Characterization of ultrashort vector pulses from a single amplitude swing measurement," Opt Express **32**, 10862–10873 (2024).
31. P. Schlup, O. Masihzadeh, L. Xu, R. Trebino, and R. A. Bartels, "Tomographic retrieval of the polarization state of an ultrafast laser pulse," Opt Lett **33**, 267–269 (2008).
32. B. Alonso and Í. Sola, "Measurement of Ultrashort Vector Pulses From Polarization Gates by In-Line, Single-Channel Spectral Interferometry," IEEE Journal of Selected Topics in Quantum Electronics **25**, 8900307 (2019).
33. D. Díaz Rivas, A.-K. Raab, C. Guo, A.-L. Viotti, I. Sytcevich, A. L'Huillier, and C. Arnold, "Measurement of ultrashort laser pulses with a time-dependent polarization state using the d-scan technique," Journal of Physics: Photonics **6**, 015003 (2024).
34. E. Szmygel, P. Béjot, A. Dubrouil, F. Billard, B. Lavorel, O. Faucher, and E. Hertz, "Characterizing ultrashort laser pulses by the rotational Doppler effect," Phys Rev A **104**, 013514 (2021).